\documentclass[12pt]{article}

\newcommand{\be}{\begin{equation}}
\newcommand{\ee}{\end{equation}}
\newcommand{\bea}{\begin{eqnarray}}
\newcommand{\eea}{\end{eqnarray}}
\newcommand{\ov}{\overline}

\newcommand{\eps}{\epsilon}
\newcommand{\ba}{\begin{array}}
\newcommand{\ea}{\end{array}}
\newcommand{\cchi}{\raisebox{2pt}{$\chi$}}

\begin{document}

\begin{flushright}
ITEP-TH-64-99\\
hep-ph/9911300
\end{flushright}

\vspace{5mm}

\centerline{\LARGE\bf Dirac operator spectral density}
\centerline{\LARGE\bf and low energy sum rules}

\vspace{5mm}

\centerline{\large K.Zyablyuk\footnote{e-mail:
    {\tt zyablyuk@heron.itep.ru}}}

\vspace{5mm}

\centerline{\it Institute of Theoretical and Experimental Physics,}
\centerline{\it B.Cheremushkinskaya 25, Moscow 117259, Russia}

\begin{abstract}
The spectral density of euclidean Dirac operator is investigated in partially quenched
QCD with arbitrary quark masses. A representation of scalar and pseudoscalar
correlators in terms of the spectral density is discussed. The spectral density, obtained from
the partially quenched chiral perturbation theory is shown to be compatible with low
energy sum rules, obtained earlier.
\end{abstract}

\section{Introduction}

One of most important characteristics in QCD is the spectral density of Dirac operator \cite{BC}:
\be
\label{sddef}
\rho(\lambda)\,=\,{1\over V}<\sum_n \delta(\lambda-\lambda_n[A])>_A \; ,
\ee
where $\lambda_n[A]$ are eingenvalues of the Dirac operator $D$ with gluon field $A$
in 4-dimensional euclidean volume $V$: $iDu_n=\lambda_n u_n$. The average
$<\ldots>_A$ denotes integration over gluon fields with Yang-Mills action and the
product of $N$ fermionic determinants $\det{(D+m_i)}$. Due to this averaging the
spectral density depends on the quark masses $m_i$.

A useful tool for the investigation of the spectral density is partially quenched QCD \cite{BG}.
In addition to $N$ usual quark with masses $m_i$ it includes one unphysical (valence) quark with
mass $m_{v1}$ and one unphysical quark of opposite statistics with mass $m_{v2}$. The generation
functional of this $SU(N+1|1)$ theory is written as follows:
\be
\label{pqgf}
Z^{pq}(m_1,\ldots,m_N;m_{v1},m_{v2})\,=\,\int [dA]\, {\det{(D +m_{v1})}\over \det{(D+m_{v2}) }} \,
\prod_{i=1}^{N} \det{(D+m_i)}\, e^{S_{YM}[A]}
\ee
Then one finds the condensate of unphysical fermionic quark for equal masses $m_{v1}$ and
$m_{v2}$:
\be
\label{sig}
\Sigma(m_v)\,=\,{1\over V} < \sum_n{1\over m_v - i\lambda_n[A]} >_A \,=\,
{1\over V} {\partial \over \partial m_{v1}}\left. \ln{Z^{pq}}(m_1,\ldots,m_N;m_{v1},m_{v2})
\right|_{m_{v1}=m_{v2}=m_v}
\ee
It depends on the masses of all quarks, but we shall write explicitly only the dependence
on the unphysical quark mass $m_v$. If we put it equal to the mass of some physical quark,
we get its chiral condensate:
\be
\Sigma(m_i)\,=\,-\,<\bar{q}_iq_i>
\ee
The spectral density is the discontinuity of $\Sigma(m_v)$ across the imaginary axis:
\be
\label{rho1}
2\pi\rho(\lambda)\,=\, \Sigma(i\lambda+0)\,-\,\Sigma(i\lambda-0) \; ,
\ee
This equation with condensate $\Sigma(m)$ given by (\ref{pqgf}, \ref{sig}) is just
trivial rewritting of the definition (\ref{sddef}) of the spectral density. Nevertheless it
allows us to compute the spectral density using well known rules and methods of
evaluating the Feynman graphs.

Indeed, the condensate of unphysical fermionic quark
$\Sigma(m_v)$ is the fermion Green function at equal coordinates.
It can be obtained from the quark propagator $S(p)$ in momentum space in $d$ dimensions
as the expansion over QCD coupling constant $\alpha_s$:
\bea
\Sigma(m) & = & {\rm tr}  \int\!{d^d p\over (2\pi)^d} \, S(p)  \nonumber \\
& = & \int\!{d^d p\over (2\pi)^d} {12 \, m \over p^2 + m^2}- 64 \pi \alpha_s m
\int\!{d^d p \, d^d k\over (2\pi)^{2d}}\, {  2(d-2)(pk) +(d-4)p^2+dm^2 \over
(p^2+m^2)^2 k^2  [(p+k)^2+m^2]}  +  O(\alpha_s^2) \nonumber \\
  & = & 12\, m^{d-1} {\Gamma(1-d/2)\over (4\pi)^{d/2}}\,-\,
 2^7 \pi \alpha_s \,m^{2d-5} {(d-1) \over (d-3)}{\Gamma^2(2-d/2)\over (4\pi)^d} \,+\, O(\alpha_s^2) \; ,
\label{cqcd1}
\eea
"tr" here is the trace over color and spinor indices, ${\rm tr} \,1=3\cdot 4$;
euclidean space is assumed. The two-loop integral
in (\ref{cqcd1}) can be computed with help of technique described, for instance, in \cite{FJJ}.
The result (\ref{cqcd1}) is divergent in four dimensions. If we put $d=4-2\eps$ and expand over
$\eps$, we get the following result:
\be
\Sigma(m)  =  {m^3 \over \tilde{\mu}{}^{2\eps}} \left\{  {\rm const} +
 {3 \over 4\pi^2} \,\ln{m^2\over \mu^2}
+{\alpha_s \over \tilde{\mu}{}^{2\eps} \pi^3 } \left[ \left( {3\over\eps} -2 \right)
\ln{m^2\over \mu^2 } - 3\,\ln\!{}^2{m^2\over \mu^2 } \right]
 \right\} \; ,
\ee
where  $\mu^2=4\pi e^{\Gamma'(1)+1} \tilde{\mu}{}^2$ according to $\ov{\rm MS}$ convention.
The terms $\sim m^3$ are not important for our purposes, since they are regular for
$m^2<0$, while logarithmic terms give us desired discontinuity. The term with
$1/\eps$ is removed by renormalization of the mass and condensate itself.
Now we may apply the equation (\ref{rho1}) in order to get the spectral density
with $\alpha_s$ correction:
\be
\label{rhoqcd}
\rho(\lambda)\,=\,{3\, |\lambda|^3 \over 4\pi^2}\left[\, 1 \,-\,{8\, \alpha_s\over\pi}
\left( \, \ln{\lambda^2\over \mu^2}\,+\,{1\over 3}\, \right)\, \right]
\ee
Notice, that the spectral density in QCD depends on the quark masses only starting from
$\alpha_s^2$ level, where the loops with usual quarks appear.

The equation (\ref{rhoqcd}) works in the QCD range of validity,
i.e. for $|\lambda|\gg \Lambda_{QCD}$. In the region $|\lambda|\sim \Lambda_{QCD}$
the formula (\ref{pqgf}) is useless, since nonperturbative effects become important.
Some attempts to extract the behavior of the spectral density in the region
$|\lambda|\ll \Lambda_{QCD}$
from low energy sum rules were undertaken in \cite{SS}, \cite{S2}, \cite{G}.
Two-point scalar correlators, known from the chiral perturbation theory (ChPT), can be presented
as certain integrals over the spectral density. This analysis demonstrates linear behavior
$\rho(\lambda) \sim (N^2-4)|\lambda|$ for $|\lambda| \gg m$.

The spectral density  has been obtained in \cite{OTV} within 1-loop partially quenched
chiral perturbation theory  \cite{BG} with $N$ quarks of equal mass. This result correctly reproduces
the 3-point scalar correlator, obtained in \cite{SS}.

In section 2 of this paper the result of \cite{OTV} is extended to the case of arbitrary quark masses.
We  take into account complete 1-loop generation functional of partially quenched ChPT
with $O(p^4)$ terms, which absorb divergencies. In section 3 the scalar and pseudoscalar
correlators are analyzed with help of spectral density representation. Many of them can
be reduced to the condensate $\Sigma(m_v)$ (and its derivatives) and to the correlators of
topological charge densities. Our result is shown to be in agreement with the sum rules, obtained
in \cite{S2}, \cite{G}.

\section{Spectral density in partially quenched ChPT}

At low energies the chiral perturbation theory \cite{GL1}, \cite{GL2} is a good replacement
of QCD. Its degrees of freedom are $N^2-1$ mesons instead of $N$ quarks. In analogy with
this construction one supposes that the low energy approximation of the partially quenched
QCD (\ref{pqgf}) is partially quenched ChPT with chiral symmetry $SU(N+1|1)$ \cite{BG}.
The effective degrees of freedom are mesons $\phi_{ij}$ of the same statistics as the product
of relevant quarks $\bar{q}_iq_j$.

The generalization of the $SU(N)$ chiral lagrangian to the partially quenched case
is obvious: everywhere the flavor trace should be replaced with supertrace. Nevertheless,
it is simpler to consider the $U(N+1|1)$ theory with super-$\eta'$ instead of pure $SU(N+1|1)$.
In this case there is no need to impose constraint ${\rm Str}\,\phi=0$ and one can consider
diagonal fields $\phi_{ii}$ as independent variables. In the final results one can take the
limit of heavy super-$\eta'$ which corresponds to the $SU(N+1|1)$ case.

The following terms of the chiral lagrangian are responsible for the chiral condensate:
\be
\label{pqch}
L=-V_0(\phi_0)+V_1(\phi_0){\rm Str}(\partial_\mu U^\dagger\partial^\mu U) +
2\,{\rm Re\, Str}\left[V_2(\phi_0)mU\right] + V_3(\phi_0) \,\partial_\mu \phi_0\partial^\mu \phi_0
\ee
where $U$ is unitary matrix of meson fields:
$$
U\,=\,\exp\left({i\sqrt{2}\phi\over F}\right) \; , \quad
\phi_0\,=\, {\rm Str}\, \phi\, =\, \sum_{i=1}^{N+2}\,\eps_i\,\phi_{ii} \; , \quad
\eps_i= \left\{ \ba{cl} +1 & i=1,\ldots, N+1 \\
 -1 & i=N+2\ea \right. \; ,
$$
$m={\rm diag}(m_1,\ldots, m_N, m_{v1}; m_{v2})$ is the quark mass matrix, $F$ is pion decay
constant. The freedom of field redefinition $\phi_{ij}\to \phi_{ij}+\delta_{ij}f(\phi_0)$ can be
used either to vanish $V_3$ or to make $V_2$ real. We shall use the last choice,
since in this case the calculation is simpler. Expanding  the lagrangian
(\ref{pqch}) up to quadratic terms, one needs the following constants:
$$
V_1(0)\,=\,{F^2\over 4} \; , \qquad
V_2(0)\,=\,{BF^2\over 2} \; , \qquad
V_0''(0)\,=\,\mu^2 \; , \qquad
V_3(0)\,=\,{\alpha\over 2} \; .
$$
There is also term $V_2''(0)\,{\rm Str}\,m\, \phi_0^{\,2}$. It belongs to the same class, as the terms
of order $O(p^4)$, so the constant $V_2''(0)$ is supposed to be small.
Since it contributes to the condensate only at the loop level, we shall ignore it.

The derivation of the one-loop chiral condensate (\ref{sig}) is standard. The propagator of
the diagonal mesons $\phi_{ii}$ was found in \cite{BG}:
\be
G_{ij}\,=\,{\delta_{ij}\,\eps_i \over p^2-M^2_i}\,-\,
{1\over (p^2-M_i^2)(p^2-M^2_j)}\left(
{1\over \alpha p^2 - \mu^2}+\sum_k{\eps_k\over p^2-M_k^2}\right)^{-1}
\ee
where $M_i^2=2Bm_i$. Later we consider only the limit $\mu^2\to \infty$. In this case the super-$\eta'$
decouples leaving us with pure $SU(N+1|1)$ theory.
In euclidean space $p^2$ changes the sign and the chiral condensate (\ref{sig})
can be written as
\be
\label{sig1}
\Sigma(m_v)\,=\,BF^2\,-\,B\sum_p\left[\, \sum_{i=1}^N \,{1\over p^2+B(m_i+m_v)}\,
- \,{1\over (p^2+2Bm_v)^2  }\left(\sum_{k=1}^N{1\over p^2+2Bm_k}\right)\!\raisebox{4mm}{${}^{-1}$}
\right]
\ee
When the volume goes to infinity the sum over momenta in (\ref{sig1}) should be replaced with
the integral over $d^4p/(2\pi)^4$. The integral is quadratically
divergent, so we shall use dimensional regularization.  We separate divergent terms according
to usual conventions of ChPT \cite{GL2}:
\be
\label{sig2}
\Sigma(m_v)\,=\,\Sigma^r(m_v)\,-\,2B^2 c \,\sum_{i=1}^N \left[\,
\left(\,1+{2\over N^2}\,\right) m_i \,+
\left(\,1-{4\over N^2}\,\right) m_v\,\right] \; ,
\ee
$$
{\rm where } \qquad
c ={1\over 16\pi^2}\mu^{d-4}\left[ {1\over d-4}
-{1\over 2}\left(\Gamma'(1) + \ln{4\pi} +1 \right)\right] \; ,
$$
$\Sigma^r(m_v)$ is regular in $d=4$ part. The divergencies are absorbed by the coupling constants
of the $O(p^4)$ lagrangian. The terms without derivatives of this lagrangian:
\bea
L^{(4)}& =&L_6\,{\rm Str}\,(\cchi^+U + U^+\cchi)^2\,
+\,L_7\,{\rm Str}\,(\cchi^+U - U^+\cchi)^2
\nonumber   \\
\label{l4}
& &+\,L_8\,{\rm Str}\,(\cchi^+U\cchi^+U + U^+\cchi U^+\cchi)
\,+\,H_2\,{\rm Str}\,(\cchi^+\cchi)   \; ,
\eea
where $\cchi=2Bm$. Adding the contribution of these terms to (\ref{sig2}), we get the final result
for the one-loop chiral condensate:
\bea
\Sigma(m_v) & = & BF^2+{B^2\over 16\pi^2}\Biggl\{
2\int\limits_0^\infty {x\,dx\over (x+m_v)^2} \left[\left(
\sum_{i=1}^N{1\over x+m_i}\right)\!\raisebox{4mm}{${}^{-1}$}
-{1\over N^2}\sum_{i=1}^N(x+m_i) \right] \nonumber \\
 & &
\left.
-\sum_{i=1}^N\left[(m_i+m_v)\ln{B(m_i+m_v)\over \mu^2}+{2\over N^2}
(m_i-2m_v) \ln{2Bm_v\over \mu^2}+{2\over N^2}(m_i-m_v)\right]
 \right\} \nonumber \\
 & &
\label{sigf}
  +\,8 B^2\left[\,
4 L_6^r(\mu) \,\sum_{i=1}^N m_i \,+
\left(\, 2L_8^r(\mu) \,+ \,H_2^r(\mu)\,\right) m_v \,\right]
\eea
where $L^r(\mu)$, $H^r(\mu)$ are renormalized constants which depend on the normalization
scale $\mu$ \cite{GL2}:
\be
L_6^r(\mu)=L_6-{N^2+2\over 16 N^2} c  \; , \quad
L_8^r(\mu)=L_8-{N^2-4\over 16 N} c \; , \quad
H_2^r(\mu)=H_2-{N^2-4\over 8 N}  c \; .
\ee
In the integral we subtracted two terms, divergent at $x=p^2/2B\to\infty$, so the total
integral is finite.
Due to boson-fermion loop cancellation the couplings of $SU(N+1|1)$ partially quenched ChPT
are renormalized exactly in the same way as the ones of usual $SU(N)$ theory.
Obviously, the total result (\ref{sigf}) does not depend on $\mu$.

Now one  can compute the discontinuity of  the condensate $\Sigma(m_v)$ across the
imaginary axis according to (\ref{rho1}) to find the spectral density. Nevertheless,
the r.h.s. of the equation (\ref{sigf}), as it stands, does not have appropriate analytical
properties for complex $m_v$. Indeed, according to Banks-Casher relation
the condensate must have the form $\Sigma(m)=mf(m^2)$, where the
function $f(m^2)$ has a cut in $m^2$ plane along half-axis $m^2<0$. The r.h.s. of (\ref{sigf})
does not satisfy this requirement.
Consequently, in order to get the spectral density from (\ref{sigf}), the formula (\ref{rho1})
should be applied with the following modification \cite{OTV}:
\be
\label{rhod2}
2\pi\,\rho(\lambda)\,=\,{\rm Disc}\,
{ m_v \over \sqrt{m_v^2}}\,\Sigma (\sqrt{m_v^2})\vert_{m_v=i\lambda}\,=
\,\Sigma(i\lambda)\,+\,\Sigma(-i\lambda)
\ee
Notice, that in QCD the equations (\ref{rho1}) and (\ref{rhod2}) give the same result (\ref{rhoqcd}).

Final result for the Dirac operator spectral density:
\bea
\rho(\lambda) & = & {BF^2\over \pi}\,+\,{B^2\over 32\pi^3}\,\Biggl\{
\,4\int\limits_0^\infty {x(x^2-\lambda^2)\over (x^2+\lambda^2)^2}
\left[\left(
\sum_{i=1}^N{1\over x+m_i}\right)\!\raisebox{4mm}{${}^{-1}$}
-\,{1\over N^2}\sum_{i=1}^N(\,x+m_i)\, \right]dx  \nonumber \\
 & &
-\,\sum_{i=1}^N\left[\, m_i \ln{B^2(\lambda^2+m_i^2)\over \mu^4}
\,-\,2\lambda\, {\rm arctg}\,{\lambda\over m_i}\, +\,
{4\over N^2}\,m_i \ln{2B|\lambda|\over \mu^2} +\,{4\pi\over N^2}\,|\lambda|\, \right. \nonumber \\
 & &
\label{rho2}
 +\, {4\over N^2} \,m_i\, -\,(32\pi)^2L_6^r(\mu)\,m_i\,\Biggr] \Biggr\}
\eea
In particular case of equal quark masses $m_i$, the integral in (\ref{rho2}) disappears
and our result coincides with eq. (83) of ref. \cite{OTV} up to the terms, linear in
mass. In our case there is no ambiguity among these terms, since we have included all
the $O(p^4)$ contributions.

In deriving the eq. (\ref{rho2}) we neglected the contribution of higher loops. Each diagram
in ChPT brings extra factor $F^{-2}$ together with loop factor $(4\pi)^{-2}$. Consequently
the expression (\ref{rho2}) for the spectral density is valid for $|\lambda| , \, m_i \ll 4\pi F$,
which is typical expansion parameter in ChPT.

Two flavor case deserves special attention. For $N=2$ there appears extra term of order
$O(p^4)$ in the chiral lagrangian:
\bea
L^{(2)}& =&{l_3\over 16}\,{\rm Str}\,(\cchi^+U + U^+\cchi)^2\,
-\,{l_7\over 16}\,{\rm Str}\,(\cchi^+U - U^+\cchi)^2
\nonumber   \\
& &+\,{1\over 4}\,(h_1+h_3)\,{\rm Str}\,(\cchi^+\cchi)\,+
\,{1\over 4}\,(h_1-h_3)\,{\rm Re}\,{\rm Sdet}\,{\cchi} \; ,
\nonumber
\eea
the couplings $l_i$, $h_i$ are the constants of the two flavor chiral lagrangian \cite{GL1}.
In this case the last line of eq. (\ref{sigf}) should be replaced with:
\be
\label{sigfn2}
\Sigma(m_v)\, = \,\ldots \,
 + \,B^2\left[ \, 2l_3(\mu)\,(m_1+m_2)\,+\,2(h_1+h_3)m_v\,+\,(h_1-h_3){m_1 m_2\over m_v}\,\right]
\ee
The spectral density for $N=2$ is given by (\ref{rho2}) with obvious replacement
$L_6\Rightarrow l_3/16$. Again, the high energy constants $h_i$ do not contribute to the
spectral density.

Few steps further can be done to evaluate the integral in (\ref{rho2}). One may notice, that
the expression inside this integral can be written in the following form:
\be
\label{sumrep}
\left(\sum_{i=1}^N{1\over x+m_i}\right)\!\raisebox{4mm}{${}^{-1}$} =
\,{1\over N^2}\sum_{i=1}^N(\,x+m_i)\,+\,\sum_{j=1}^{N-1} \,{a_j\over x+ \tilde{m}_j}
\ee
The masses $\tilde{m}_j$ are the roots (with opposite sign) of the polinomial, which stands
in denominator of the l.h.s. in equation (\ref{sumrep}). The constants $a_j$ have dimension
(mass)${}^2$; they vanish when all $m_i$ are equal. In particular case of two quarks:
$$
N=2: \qquad \qquad \tilde{m}_1\,=\,{1\over 2}\,(m_1+m_2) \; , \qquad a_1\,=\,-\,{1\over 8}\,
(m_1-m_2)^2
$$
For $N=3$ the constants $\tilde{m}_j$ and $a_j$ also can be found in explicit form,
but we shall not write them here.
Knowing these constants, the integral in (\ref{rho2}) can be easily computed:
$$
\int\limits_0^\infty {x(x^2-\lambda^2)\over (x^2+\lambda^2)^2}
\left[\left(
\sum_{i=1}^N{1\over x+m_i}\right)\!\raisebox{4mm}{${}^{-1}$}
-\,{1\over N^2}\sum_{i=1}^N(\,x+m_i)\, \right]dx \,=
$$
\be
=\,\sum_{j=1}^{N-1} \,{a_j \tilde{m}_j \over (\lambda^2+\tilde{m}_j^2)^2} \left[ \,
(\lambda^2-\tilde{m}_j^2)\,\ln{|\lambda|\over \tilde{m}_j}\,- \,\lambda^2 \,-\,\tilde{m}_j^2 \,+
\,\pi|\lambda|\tilde{m}_j\, \right]
\ee

\section{Low energy sum rules}

The low energy sum rules are formulated for the correlators
of charge densities which can be written as certain integrals
over the spectral density. In particular, the
two point correlators of scalar and pseudoscalar charge densities
$$
S^{ij}\,=\,\bar{q}{}^i q^j \; , \qquad P^{ij}\,=\,i\bar{q}{}^i\gamma^5 q^j
$$
were expressed in terms of the spectral density in \cite{S2}.
This technique can be extended to the case of different quark masses:
\bea
\label{2sc1}
i\!\int\!dx <S^{ij}(x)S^{kl}(0)^\dagger> & = & 2\,\delta^{ik}\delta^{jl}\!\int\limits_0^\infty\!
{(\lambda^2-m_i m_j)\,\rho(\lambda)\over (\lambda^2+m_i^2)(\lambda^2+m_j^2)} \,d\lambda
\; , \quad  i\not= j, \; k\not= l ;   \quad \\
\label{2pc1}
i\!\int\!dx <P^{ij}(x)P^{kl}(0)^\dagger> & = & 2\,\delta^{ik}\delta^{jl}\!\int\limits_0^\infty\!
{(\lambda^2+m_i m_j)\,\rho(\lambda)\over (\lambda^2+m_i^2)(\lambda^2+m_j^2)}\, d\lambda
\,+\,\delta^{ij}\delta^{kl}<\nu^2> \quad
\eea
where $<\nu^2>$ is the topological susceptibility:
\be
<\nu^2>\,=\,i\!\int\! dx <\omega(x)\,\omega(0)> \; , \qquad
\omega={1\over 16\pi^2}\, {\rm tr}_c(G_{\mu\nu}\tilde{G}{}^{\mu\nu})
\ee
Correlators of diagonal scalar charges include additional contributions
of disconnected graphs. The insertion of the operator $-i\int dx S^{ii}(x)$ corresponds
to the differentiation over the mass $m_i$. In particular
\be
\label{2scd}
i\!\int\!dx <S^{ii}(x)S^{jj}(0)^\dagger> \,  = \, {\partial \Sigma(m_j)\over \partial m_i}
\ee
Since the condensate is itself the derivative, l.h.s. is symmetric over $ij$.

There is no need to compute the integrals in (\ref{2sc1},\ref{2pc1}), since they can be expressed
in terms of the chiral condensate:
\bea
i\!\int\! dx <S^{ij}(x)S^{kl}(0)^\dagger> & = & \delta^{ik}\delta^{jl}\,
{\Sigma(m_i)-\Sigma(m_j)\over m_i-m_j} \; , \quad  i\not= j, \; k\not= l ;  \label{2sc2} \\
i\!\int\! dx <P^{ij}(x)P^{kl}(0)^\dagger> & = & \delta^{ik}\delta^{jl}\,
{\Sigma(m_i)+\Sigma(m_j)\over m_i+m_j}\,+\,\delta^{ij}\delta^{kl}<\nu^2> \label{2pc2}
\eea
These equations directly follow from the limit $q^\mu\to 0$ in the appropriate vector/axial correlators
$q^\mu q^\nu\int e^{iqx}dx<V_\mu(x) V_\nu(0)>$. Nevertheless, the equation (\ref{2sc2}) within
the framework of the partially quenched theory can be also applied if some quarks
have equal masses: the ratio should be replaced with derivative. The result of this limit
$m_i\to m_j$ is however not the same as the l.h.s. of (\ref{2scd}) due to different order
of operations: one should compute the derivative $\partial \Sigma(m_v)/\partial m_v$  at first and
substitute $m_v= m_i$ after then.

In the same way other n-point off-diagonal correlators can be expressed in terms of
the chiral condensate. In particular
$$
i^{n-1}\!\int\! dx_1 \ldots dx_{n-1} <S^{i_1j_1}(x_1)\ldots S^{i_{n-1}j_{n-1}}(x_{n-1})S^{i_nj_n}(0)>\,=
\hspace{39mm}
$$
\be
\label{nsnd}
=\left( \delta^{i_1j_2}\ldots\delta^{i_{n-1}j_n} \delta^{i_nj_1} + perm(j) \right)
\sum_{a=1}^n {\Sigma(m_{i_a})\over \prod\limits_{b=1, \, b\not= a}^n (m_{i_b}-m_{i_a})} \; , \qquad
i_c\not= j_c \;\, \mbox{for any} \;\, c \, .
\ee
where $perm(j)$ denotes all permutations of indices $j$.

A large set of sum rules can be formulated using spectral density representation.
Below we consider the simplest realistic model with $N=2$ flavors of equal mass $m_u=m_d=m$.
The charges are combined in triplet $S^a=\bar{q}\sigma^a q$ and singlet $S^0=\bar{q}q$.
The following sum rules have been formulated earlier \cite{S2}, \cite{G}:
\bea
i\!\int\! dx <S^a(x)S^b(0)-\delta^{ab}P^0(x)P^0(0)> & = & 2\,\delta^{ab}
\left( \left.{\partial \Sigma(m_v)\over \partial m_v}\right|_{m_v=m}\!-
{\Sigma(m)\over m} -2<\nu^2> \right)    \nonumber \\
\label{sr1}
& = &-\delta^{ab} 8B^2 l_7 \; ; \\      & &  \nonumber \\
i\!\int\! dx <\delta^{ab}S^0(x)S^0(0)-P^a(x)P^b(0)> & = & 2\,\delta^{ab}
\left( {\partial \Sigma(m)\over \partial m}- {\Sigma(m)\over m} \right) \nonumber \\
\label{sr2}
& = &\delta^{ab}\left( -{2BF^2\over m} - {3B^2\over 8\pi^2} \right) \; .
 \eea
We have expressed the integrals over the spectral density in terms of the condensate
according to (\ref{2scd}-\ref{2pc2}). Equation (\ref{sr2}) is satisfied for $\Sigma(m_v)$ given
by (\ref{sigf}). The sum rule (\ref{sr1}) gives the following result for the topological susceptibility:
\be
<\nu^2>\,=\,{1\over 2}\,BF^2m\,+\,B^2 m^2\left(\,2\,l_3^r(\mu)\,-\,2\,l_7\,-
\,{3\over 32\pi^2}\ln{M_\pi^2\over \mu^2}\,\right)
\ee
This result can be obtained directly from the two flavor chiral lagrangian \cite{GL1},
which verifies the partially quenched approach used here.

\section*{Acknowledgements}

Author thanks A.Smilga and A.Gorsky for suggesting the problem
and discussions.

\end{document}